\documentstyle[twoside]{article}

\catcode`\@=11
\long\def\@makefntext#1{
\protect\noindent \hbox to 3.2pt {\hskip-.9pt  
$^{{\eightrm\@thefnmark}}$\hfil}#1\hfill}		

\def\@makefnmark{\hbox to 0pt{$^{\@thefnmark}$\hss}}	
	
\def\ps@myheadings{\let\@mkboth\@gobbletwo
\def\@oddhead{\hbox{}
\rightmark\hfil\eightrm\thepage}   
\def\@oddfoot{}\def\@evenhead{\eightrm\thepage\hfil
\leftmark\hbox{}}\def\@evenfoot{}
\def\sectionmark##1{}\def\subsectionmark##1{}}

\oddsidemargin=\evensidemargin
\newcounter{sectionc}\newcounter{subsectionc}\newcounter{subsubsectionc}
\renewcommand{\section}[1] {\vspace{12pt}\addtocounter{sectionc}{1} 
\setcounter{subsectionc}{0}\setcounter{subsubsectionc}{0}\noindent 
	{\tenbf\thesectionc. #1}\par\vspace{5pt}}
\renewcommand{\subsection}[1] {\vspace{12pt}\addtocounter{subsectionc}{1} 
	\setcounter{subsubsectionc}{0}\noindent 
	{\bf\thesectionc.\thesubsectionc. {\kern1pt \bfit #1}}\par\vspace{5pt}}
\renewcommand{\subsubsection}[1] {\vspace{12pt}\addtocounter{subsubsectionc}{1}
	\noindent{\tenrm\thesectionc.\thesubsectionc.\thesubsubsectionc.
	{\kern1pt \tenit #1}}\par\vspace{5pt}}
\newcommand{\nonumsection}[1] {\vspace{12pt}\noindent{\tenbf #1}
	\par\vspace{5pt}}

\newcounter{appendixc}
\newcounter{subappendixc}[appendixc]
\newcounter{subsubappendixc}[subappendixc]
\renewcommand{\thesubappendixc}{\Alph{appendixc}.\arabic{subappendixc}}
\renewcommand{\thesubsubappendixc}
	{\Alph{appendixc}.\arabic{subappendixc}.\arabic{subsubappendixc}}

\renewcommand{\appendix}[1] {\vspace{12pt}
        \refstepcounter{appendixc}
        \setcounter{figure}{0}
        \setcounter{table}{0}
        \setcounter{lemma}{0}
        \setcounter{theorem}{0}
        \setcounter{corollary}{0}
        \setcounter{definition}{0}
        \setcounter{equation}{0}
        \renewcommand{\thefigure}{\Alph{appendixc}.\arabic{figure}}
        \renewcommand{\thetable}{\Alph{appendixc}.\arabic{table}}
        \renewcommand{\theappendixc}{\Alph{appendixc}}
        \renewcommand{\thelemma}{\Alph{appendixc}.\arabic{lemma}}
        \renewcommand{\thetheorem}{\Alph{appendixc}.\arabic{theorem}}
        \renewcommand{\thedefinition}{\Alph{appendixc}.\arabic{definition}}
        \renewcommand{\thecorollary}{\Alph{appendixc}.\arabic{corollary}}
        \renewcommand{\theequation}{\Alph{appendixc}.\arabic{equation}}
        \noindent{\tenbf Appendix \theappendixc #1}\par\vspace{5pt}}
\newcommand{\subappendix}[1] {\vspace{12pt}
        \refstepcounter{subappendixc}
        \noindent{\bf Appendix \thesubappendixc. {\kern1pt \bfit #1}}
	\par\vspace{5pt}}
\newcommand{\subsubappendix}[1] {\vspace{12pt}
        \refstepcounter{subsubappendixc}
        \noindent{\rm Appendix \thesubsubappendixc. {\kern1pt \tenit #1}}
	\par\vspace{5pt}}

\newcommand{\textlineskip}{\baselineskip=13pt}
\newcommand{\smalllineskip}{\baselineskip=10pt}

\def\eightcirc{
\begin{picture}(0,0)
\put(4.4,1.8){\circle{6.5}}
\end{picture}}
\def\eightcopyright{\eightcirc\kern2.7pt\hbox{\eightrm c}} 

\newcommand{\copyrightheading}[1]
	{\vspace*{-2.5cm}\smalllineskip{\flushleft
	{\footnotesize International Journal of Modern Physics A, #1}\\
	{\footnotesize $\eightcopyright$\, World Scientific Publishing
	 Company}\\
	 }}

\def\abstracts#1#2#3{{
	\centering{\begin{minipage}{4.5in}\baselineskip=10pt\footnotesize
	\parindent=0pt #1\par 
	\parindent=15pt #2\par
	\parindent=15pt #3
	\end{minipage}}\par}} 

\renewenvironment{thebibliography}[1]
	{\frenchspacing
	 \ninerm\baselineskip=11pt
	 \begin{list}{\arabic{enumi}.}
	{\usecounter{enumi}\setlength{\parsep}{0pt}
	 \setlength{\leftmargin 12.7pt}{\rightmargin 0pt} 
	 \setlength{\itemsep}{0pt} \settowidth
	{\labelwidth}{#1.}\sloppy}}{\end{list}}

\newcounter{itemlistc}
\newcounter{romanlistc}
\newcounter{alphlistc}
\newcounter{arabiclistc}

\newcommand{\fcaption}[1]{
        \refstepcounter{figure}
        \setbox\@tempboxa = \hbox{\footnotesize Fig.~\thefigure. #1}
        \ifdim \wd\@tempboxa > 5in
           {\begin{center}
        \parbox{5in}{\footnotesize\smalllineskip Fig.~\thefigure. #1}
            \end{center}}
        \else
             {\begin{center}
             {\footnotesize Fig.~\thefigure. #1}
              \end{center}}
        \fi}

\newcommand{\tcaption}[1]{
        \refstepcounter{table}
        \setbox\@tempboxa = \hbox{\footnotesize Table~\thetable. #1}
        \ifdim \wd\@tempboxa > 5in
           {\begin{center}
        \parbox{5in}{\footnotesize\smalllineskip Table~\thetable. #1}
            \end{center}}
        \else
             {\begin{center}
             {\footnotesize Table~\thetable. #1}
              \end{center}}
        \fi}

\def\@citex[#1]#2{\if@filesw\immediate\write\@auxout
	{\string\citation{#2}}\fi
\def\@citea{}\@cite{\@for\@citeb:=#2\do
	{\@citea\def\@citea{,}\@ifundefined
	{b@\@citeb}{{\bf ?}\@warning
	{Citation `\@citeb' on page \thepage \space undefined}}
	{\csname b@\@citeb\endcsname}}}{#1}}

\newif\if@cghi
\def\cite{\@cghitrue\@ifnextchar [{\@tempswatrue
	\@citex}{\@tempswafalse\@citex[]}}
\def\citelow{\@cghifalse\@ifnextchar [{\@tempswatrue
	\@citex}{\@tempswafalse\@citex[]}}
\def\@cite#1#2{{$\null^{#1}$\if@tempswa\typeout
	{IJCGA warning: optional citation argument 
	ignored: `#2'} \fi}}

\def\pmb#1{\setbox0=\hbox{#1}
	\kern-.025em\copy0\kern-\wd0
	\kern.05em\copy0\kern-\wd0
	\kern-.025em\raise.0433em\box0}

\def\fnt#1#2{\footnotetext{\kern-.3em
	{$^{\mbox{\scriptsize #1}}$}{#2}}}

\def\fpage#1{\begingroup
\voffset=.3in
\thispagestyle{empty}\begin{table}[b]\centerline{\footnotesize #1}
	\end{table}\endgroup}

\def\runninghead#1#2{\pagestyle{myheadings}
\markboth{{\protect\footnotesize\it{\quad #1}}\hfill}
{\hfill{\protect\footnotesize\it{#2\quad}}}}
\font\tenrm=cmr10
\font\tenit=cmti10 
\font\tenbf=cmbx10
\font\bfit=cmbxti10 at 10pt
\font\ninerm=cmr9

\font\eightrm=cmr8

\def\qed{\hbox{${\vcenter{\vbox{			
   \hrule height 0.4pt\hbox{\vrule width 0.4pt height 6pt
   \kern5pt\vrule width 0.4pt}\hrule height 0.4pt}}}$}}

\begin{document}

\runninghead{Low Dirac Eigenmodes $\ldots$} {Low Dirac Eigenmodes $\ldots$}

\normalsize\textlineskip
\thispagestyle{empty}
\setcounter{page}{1}

\copyrightheading{}			

\vspace*{0.88truein}

\fpage{1}
\centerline{\bf LOW DIRAC EIGENMODES AND THE TOPOLOGICAL AND}
\vspace*{0.035truein}
\centerline{\bf CHIRAL STRUCTURE OF THE QCD VACUUM\footnote{Talk
presented at the DPF2000 Conference .}}
\vspace*{0.37truein}
\centerline{\footnotesize H. B. THACKER}
\vspace*{0.015truein}
\centerline{\footnotesize\it Department of Physics,
University of Virginia}
\baselineskip=10pt

\vspace*{0.21truein}
\abstracts{Several lattice calculations which probe the chiral and topological 
structure of QCD are discussed. The results focus attention on the low-lying eigenmodes
of the Dirac operator in typical gauge field configurations.}{}{}


\vspace*{1pt}\textlineskip	
\section{Eigenmode expansion of quark propagators}	
\vspace*{-0.5pt}
\noindent

Standard methods in lattice QCD consist of calculating various hadronic amplitudes
by tying together appropriate combinations of quark propagators calculated on an
ensemble of Monte Carlo generated gauge fields. A number of recent lattice QCD calculations have focused increasing attention
on the role of low-lying eigenmodes of the Dirac operator in typical
background gauge fields. 
The decomposition of quark propagators into a 
sum over eigenmodes provides some very useful insight into the structure of low-enery QCD, particularly in exposing connections between quark propagation
and chiral dynamics. 

\section{Lattice Results}
\noindent

I will breifly
describe several recent and ongoing investigations which employ lattice calculations
to address issues of chiral symmetry, gauge topology, and the structure and
implications of low Dirac eigenmodes.
All of these are discussed in more detail in the cited references.  

\subsection{Exceptional configurations and pole-shifting}
\noindent
The cause of the ``exceptional configuration'' problem for quenched QCD calculations
at light quark mass has been found.\cite{MQA} It arises from
the presence of a small number of exactly real eigenmodes which should be at 
zero quark mass (critical hopping parameter) but which, on some
configurations, have migrated to positive quark mass due to the lattice chiral
symmetry breaking of the Wilson mass term. The cure consists of shifting these
positive mass eigenmodes to zero mass where they belong. Meson and hairpin
correlators are then calculated with good statistics down to pion masses
of $200$ MeV or less. 

\subsection{The $\eta'$ mass}
\noindent
Using improved ``pole-shifted'' quark propagators, the pseudoscalar hairpin 
(loop-loop) correlator
has been calculated for quark masses ranging from approximately the strange quark
mass down to very light quark mass, corresponding to a pion mass less than $200$
MeV.\cite{chlogs} The time-dependence 
of the quenched hairpin correlator is very well described by the double-Goldstone pole
form expected from a chiral Lagrangian description in which the $q\bar{q}$ annihilation-creation
process is treated as an $\eta'$ mass insertion. 
At $\beta=5.7$, the anomaly contribution to the $\eta'$ mass is $m_0=464(18)$ for unimproved
Wilson quarks, and $685(32)$ MeV for clover-improved quarks, using $C_{sw}=1.57$,
and a lattice spacing of $a^{-1}=1.18$ GeV. 

\subsection{Topological susceptibility of quenched QCD}
\noindent
The integrated anomaly or fermionic method for determining topological susceptibility,
originally suggested by Smit and Vink, has been applied to the same ensembles
of quenched gauge configurations used to study the $\eta'$ mass.\cite{chlogs} The winding
number $\nu$ of each configuration is determined by integrating the pseudoscalar charge
and using the integrated axial anomaly, $\int\bar{\psi}\gamma^5\psi d^4x = i\nu/m$.
The result obtained (with clover improved fermions) 
for the topological susceptibility is
$\chi_t=(188(4)\;MeV)^4$. This is in good agreement with a phenomelogical estimate 
based on the $\eta'$ mass and the Witten-Veneziano formula.

\subsection{Quenched chiral logarithms}
\noindent
Anomalous chiral effects of the quenched $\eta'$ propagator, originally discussed by Sharpe,
Bernard, and Golterman \cite{SB&G} have been confirmed by lattice results.\cite{chlogs} 
The predicted anomalous power law behavior of
both the pion mass as a function of quark mass and of the pseudoscalar decay constant as
a function of $m_{\pi}^2$ are observed. Also as predicted, the axial vector decay constant
is found to have smooth analytic behavior in the chiral limit. All the observed quenched
chiral log effects are consistent with a value of the 
QCL parameter of $\delta=.065(13)$ for
clover improved quarks.

\subsection{Ghost states in the scalar valence propagator}
\noindent
A study of the scalar, isovector meson propagator in quenched QCD reveals a striking
behavior which is another QCL effect of the quenched
$\eta'$ propagator.\cite{ghosts}  As the quark mass becomes light,
the scalar meson propagator develops a long-range tail which falls off rougly like
$\exp(-2m_{\pi}t)$ and {\it has the wrong sign}, i.e. the long range tail has a sign
opposite to that required by spectral positivity. The results agree in detail with
the assumption that this tail represents the contribution
of a chiral loop diagram containing an $\eta'$-$\pi$ intermediate state, which is
light and effectively of negative spectral weight in the quenched theory. In quenched
chiral perturbation theory,\cite{SB&G} the tail arises from a ghost fermion-antifermion loop. 

\subsection{The truncated determinant algorithm}
\noindent
The truncated determinant algorithm (TDA)\cite{TDA} is an approach to full QCD simulations that
begins by splitting the closed quark-loop determinant into a product over low Dirac eigenvalues
times a product over high eigenvalues $\Delta = \Delta_l\Delta_h$. The separation
is typically made at about 600 to 800 MeV. The low end of the determinant $\Delta_l$
is treated exactly, using the Lanczos algorthm combined with a Metropolis accept-reject step.
The high end of the determinant $\Delta_h$
is found to be well approximated by a set of gauge
loops, which can be easily included in the Monte Carlo heat bath. 

\subsection{The origin of the OZI rule}

Using methods similar to those used to study the $\eta'$ hairpin correlator,
$q\bar{q}$ annihilation-creation processes in other spin-parity channels have been
studied.\cite{OZI} The results confirm the strong OZI suppression in the vector and axial vector
channels expected from phenomenology. OZI violation in the scalar channel is found to be
large and comparable to the pseudoscalar channel, as predicted from quark model
arguments by Isgur and Geiger.\cite{IG} The OZI rule is found to be fundamentally a statement about
the spin-parity structure of low-lying Dirac eigenmodes.

\subsection{Evidence against instantons in QCD}

Arguments by Witten that topological charge fluctuations in the QCD vacuum come not in the
form of instantons but rather from the generically large gauge fluctuations of a confining
vacuum are strongly supported by a study of the chiral structure of low Dirac eigenmodes.\cite{noinst}

\nonumsection{Acknowledgements}
\noindent
I am grateful to my collaborators W. Bardeen, A. Duncan, E. Eichten, I. Horvath, N. Isgur,
and J. McCune. This work was supported in part by the U. S. Department of Energy 
under grant DE-FG02-97ER41027.

\nonumsection{References}

\end{document}